\documentclass{article}
\usepackage{spconf}

\usepackage{cite}
\usepackage{amsmath,amssymb,amsfonts}
\usepackage{algorithmic}
\usepackage{graphicx}
\usepackage{textcomp}
\usepackage{xcolor}
\usepackage{multicol}
\usepackage{multirow}
\usepackage{graphicx}
\usepackage{color}
\usepackage{comment}
\usepackage{booktabs}
\usepackage{multicol}
\usepackage{multirow}
\usepackage{lipsum}
\usepackage{url}
\usepackage{hyperref}

\newcommand{\absI}{\texttt{abs-mfb-nsfs}}
\newcommand{\absII}{\texttt{abs-mfb-nsf}}
\newcommand{\absIII}{\texttt{abs-mfb-nsfg}}
\newcommand{\absIV}{\texttt{abs-mfb-hfg}}
\newcommand{\sysI}{\texttt{abs-mfbf-nsfs}}
\newcommand{\sysII}{\texttt{taco-mfbf-nsfs}}
\newcommand{\sysIII}{\texttt{taco-mfb-nsfs}}
\newcommand{\sysIV}{\texttt{taco-mfb-nsf}}
\newcommand{\sysV}{\texttt{taco-mfb-nsfg}}
\newcommand{\sysVI}{\texttt{taco-mfb-hfg}}
\newcommand{\sysVII}{\texttt{trans-mfb-nsfs}}
\newcommand{\sysVIII}{\texttt{trans-mfb-nsf}}
\newcommand{\sysIX}{\texttt{trans-mfb-nsfg}}
\newcommand{\sysX}{\texttt{trans-mfb-hfg}}
\newcommand{\sysXI}{\texttt{joint-nsf}}
\newcommand{\sysXII}{\texttt{joint-nsfg}}
\newcommand{\sysXIII}{\texttt{joint-hfg}}
\newcommand{\sysXIV}{\texttt{Fluidsynth}}
\newcommand{\sysXV}{\texttt{Pianoteq}}

\def\BibTeX{{\rm B\kern-.05em{\sc i\kern-.025em b}\kern-.08em
    T\kern-.1667em\lower.7ex\hbox{E}\kern-.125emX}}

\title{
Can Knowledge of End-to-End Text-to-Speech Models Improve Neural MIDI-to-Audio Synthesis Systems?}
\name{Xuan Shi$^1$\thanks{This study is partially supported by the Japanese-French joint national project called VoicePersonae supported by JST CREST (JPMJCR18A6, JPMJCR20D3), MEXT KAKENHI Grants (21K17775, 21H04906, 21K11951), Japan, and Google AI for Japan program.}, Erica Cooper$^2$, Xin Wang$^2$, Junichi Yamagishi$^2$, Shrikanth Narayanan$^1$}
\address{$^1$University of Southern California, USA
  $^2$National Institute of Informatics, Japan}

\begin{document}

\ninept

\maketitle

\begin{abstract}
With the similarity between music and speech synthesis from symbolic input and the rapid development of text-to-speech (TTS) techniques, it is worthwhile to explore ways to improve the MIDI-to-audio performance by borrowing from TTS techniques.  In this study, we analyze the shortcomings of a TTS-based MIDI-to-audio system and improve it in terms of feature computation, model selection, and training strategy, aiming to synthesize highly natural-sounding audio.  Moreover, we conducted an extensive model evaluation through listening tests, pitch measurement, and spectrogram analysis.  This work demonstrates not only synthesis of highly natural music but offers a thorough analytical approach and useful outcomes for the community.  Our code, pre-trained models, supplementary materials, and audio samples are open sourced at \href{https://github.com/nii-yamagishilab/midi-to-audio}{https://github.com/nii-yamagishilab/midi-to-audio} \label{projweb}.

\end{abstract}


\section{Introduction}

Generating musical instrument sounds using data-driven neural synthesis approaches allows us to benefit from advances in sequence-to-sequence modeling \cite{cooper2021midi, dong2022deepperformer}, to learn from audio examples \cite{wang2019performancenet, kumar2019melgan}, and to flexibly interpolate and transform instrument sounds \cite{kim2019neural, wu2022mididdsp}. 
Compared with sample-based synthesis such as {\sysXIV} \cite{FluidSynth} and physical modeling based approaches such as {\sysXV} \cite{Pianoteq}, data-driven neural MIDI-to-audio synthesis allows us to flexibly generate new instrument sounds without having to painstakingly specify the physical parameters of every instrument by hand.

Although there are differences between musical instrument sounds and speech sounds, we showed in our past work \cite{cooper2021midi} that text-to-speech synthesis architectures composed of an acoustic model and a neural vocoder can be modified for the task of generating piano sounds from MIDI input.  This opens up many possibilities for musical sound generation that conceptually match well with tasks in speech generation, such as musical style transfer using style tokens \cite{cifka2021self, liang2017automatic}, multi-instrument modeling in the manner of multi-language multi-speaker TTS \cite{cooper2020zero, hawthorne2022multi}, and timbre transfer and interpolation based on voice conversion frameworks \cite{kim2019neural, engel2020ddsp}.  

However, while speech synthesis has reached a level of quality where it can be almost indistinguishable from human speech, we still observe a gap in quality between our synthesized audio and real piano sounds. In this paper, we address this quality gap with several proposed improvements, and we analyze our synthesized audio to identify further areas for improvement.

The first of our proposed improvements is joint training of the acoustic model and neural vocoder. In TTS, the acoustic model and neural vocoder are often separately trained, but it is also known that this results in degraded sound quality \cite{shen2018natural,kim2021conditional} since separate training may cause a mismatch due to the fact that the neural vocoder is trained using natural spectrograms, whereas generated spectrograms are input at inference time.  We hypothesize that the joint training of the two models will improve the quality of the generated audio even for the case of musical instrument sounds. 
Second, we introduce improved acoustic model and vocoder architectures in terms of training efficiency.  In our prior work we used a Tacotron 2 \cite{shen2018natural} architecture with a neural source-filter (NSF) \cite{wang2019neural} vocoder, and in this work, we explore the Transformer TTS architecture \cite{li2019neural} as well as the HiFiGAN vocoder \cite{kong2020hifi} and its combination with NSF. 
Finally,  
our prior work used the MIDI-scale spectrogram instead of the Mel spectrogram, but the number of fast Fourier transform (FFT) points was not big enough to perfectly support a MIDI-scale filter bank. We fix this issue by using a larger number of FFT points.

We evaluated our proposed improvements in a crowdsourced mean opinion score (MOS) listening test, which revealed that the improved system using the above refinements can lead to high-fidelity waveform signals and produce a higher quality of audio than the standard MIDI player. However, the results also show that there is still a gap between natural and generated audio. 
We further analyzed the objective and subjective results at the note level and identified areas for further improvement.

This paper is structured as follows: Section 2 reviews related work on MIDI-to-audio synthesis, and the proposed improvements are described in Section 3. Section 4 describes the experiments, and our findings are summarized in Section 5.

\section{Background}\label{sec:related_work}

Several deep learning-based algorithms have been proposed to generate music from musical notes. 
PerformanceNet \cite{wang2019performancenet} uses a U-Net-based ContourNet and a multi-band residual network-based TextureNet to transfer piano roll input to a spectrogram.
It then uses the Griffin-Lim algorithm \cite{griffin1984signal} to reconstruct a waveform.
Derived from DDSP \cite{engel2020ddsp}, MIDI-DDSP \cite{wu2022mididdsp} integrates three separately trained modules in a hierarchical way to control the notes, performance, and synthesis factors, respectively.  MIDI-DDSP is capable of synthesizing high-fidelity audio while keeping interactive control of the synthesis. 
Deep Performer \cite{dong2022deepperformer} explicitly models necessary factors for audio synthesis with individual models and uses a Transformer to generate a Mel-spectrogram. It then uses a separately trained HiFi-GAN to transform the Mel-spectrogram into music audio.
Our previous work \cite{cooper2021midi} observed a similarity between TTS and MIDI-to-audio synthesis and used a Tacotron-based acoustic model and an NSF-based waveform model to synthesize audio from aligned piano roll input.

\section{MIDI-to-audio synthesis system based on TTS techniques and its improvements}

\subsection{Overview of baseline MIDI-to-audio synthesis system}

The MIDI-to-audio synthesis process is similar to TTS in the sense that both convert symbolic data into audio signals. 

The input MIDI contains messages that encode the note identity, the time of note onset and offset, velocity, and other events. For processing using deep learning models, MIDI input is often represented as a piano roll, which is a sequence of multi-hot vectors, where each multi-hot vector is 128 dimensions.  Each dimension encodes the velocity for one of the 128 MIDI notes.

Using a sequence of multi-hot vectors as an input, a sequence-to-sequence model transforms it into a sequence of acoustic features that are suitable to drive a neural waveform model. In our previous work \cite{cooper2021midi}, we used a variant of Tacotron 2 with self-attention as the acoustic model \cite{yasuda2019investigation} and a 128-dimensional MIDI-scaled spectrogram \cite{cooper2021midi}, which is similar to the  Mel spectrogram but uses a filter-bank located at the center frequencies of MIDI notes. As for the neural waveform model, we used NSF\cite{wang2019neural}. 

\begin{figure}[t!]
\begin{center}
\includegraphics[width=0.9\columnwidth]{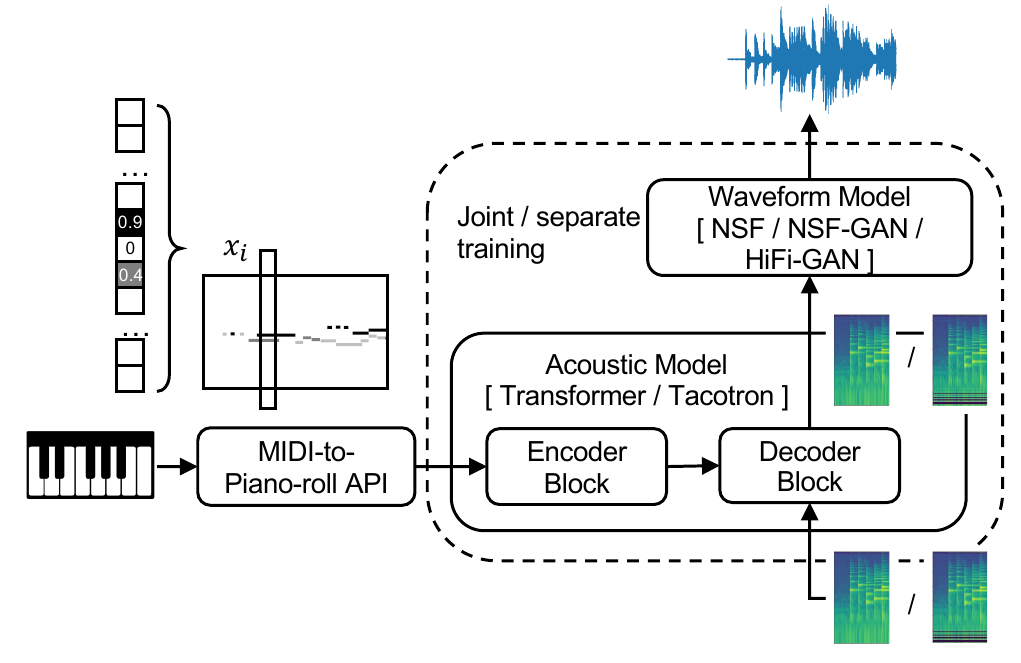}
\vspace{-5mm}
\caption{Architecture of investigated methods.}
\label{fig:app_example}
\end{center}
\vspace{-6mm}
\end{figure}

\vspace{-1mm}
\subsection{Shortcomings of baseline and proposed refinements}
\vspace{-1mm}

Our previous study showed that the generated audio samples from the baseline system with the best configuration achieved an acceptable MOS of around 3.2 \cite{cooper2021midi}, but it was significantly lower than the score of 4.0 obtained for the natural audio. One major factor we identified is the feature mismatch for the input to the neural vocoder. While the neural vocoder was trained with acoustic features extracted from natural audio as input, it took as input the acoustic features generated by the acoustic model during inference. 

A similar issue had raised concerns in the TTS community and inspired a few methods for refinement. We are motivated to analyze how well these methods can be applied to the MIDI-to-audio sound generation task.
The first improvement is joint training of the acoustic model and neural vocoder.  
This is done by concatenating the two models, conducting back-propagation from the generated audio waveform to the input MIDI sequence, and jointly optimizing both modules. By doing so, the neural vocoder takes the acoustic model's output as input both in the training and inference stage, which alleviates the feature mismatch. 

Since the two modules affect the performance of the whole system, the second refinement is to explore better architectures for each of them. In this study, we investigate the Transformer TTS architecture \footnote{We directly take the pianoroll as input of the acoustic model like \cite{cooper2021midi}, instead of pre-processing music notes into multiple embeddings in \cite{dong2022deepperformer}.} and the HiFi-GAN (and its combination with NSF) as improved acoustic model and neural vocoders, respectively. 
The reason to investigate Transformer is its superior performance in speech synthesis \cite{li2019neural} and other audio processing tasks \cite{huang2018music}. 
As for waveform models, NSF, which uses only a multi-resolution short-time Fourier transform (STFT) loss \cite{wang2019neural}, suffers from the limitations of a maximum-likelihood-based model training criterion. We expect that GAN-based waveform models such as HiFi-GAN can further improve the system performance. Meanwhile, we can combine the NSF model and the HiFi-GAN discriminators to address the limitations of NSF. We refer to this model as NSF-GAN.

Another refinement is on the acoustic features. 
The original implementation of the MIDI spectrogram uses only a 4,096-point FFT, but the width of the first few filters in the MIDI filter-bank is narrower than 2 Hz and may not cover any FFT frequency bin. Hence, as shown in the Fig.4 of \cite{cooper2021midi}, these filter outputs, or the corresponding feature dimensions of the MIDI spectrogram, become zero.
As refinement, we increase the number of FFT points by a factor of four to 16,384 when extracting the MIDI-spectrogram.

\section{Experiments}
\label{sec:experiments}

\vspace{-1mm}
\subsection{Experimental condition}
\vspace{-1mm}

\noindent
\textbf{Database}: 
We used the MAESTRO dataset \cite{hawthorne2018enabling} to explore piano audio synthesis. 
MAESTRO was collected from the international Piano-e-Competition. In addition to the concert-quality piano performances presented by virtuoso pianists, a high-precision sensor built into the piano captures the time-aligned MIDI transcription of each performance with an accuracy of 3 ms. The MAESTRO dataset has 1,276 performances totaling 198.7 hours, which were divided into training, validation, and test sets with 962, 137, and 177 performances totaling respectively 159.2, 19.4, and 20.0 hours.  

In our experiments, we followed the split configuration in \cite{hawthorne2018enabling} but resampled the audio to 24 kHz. 
We further preprocessed the data in two ways to accommodate the synthesis system.  First, we segmented each long sample into short pieces of 800 frames for training to avoid memory issues; second, we converted the MIDI input, which uses an onset/offset representation of notes, into a piano roll format to make it simpler for the model to parse the notes.    
We used the PrettyMIDI API \cite{raffel2014intuitive} to convert MIDI files into piano rolls.

\noindent\textbf{Systems}:
Experimental models are listed in the left-most column of Table~\ref{tab:midi-to-audio}. Following \cite{cooper2021midi}, we included the sample-based synthesizer {\sysXIV} and the physical model synthesizer {\sysXV} as reference. Other systems followed the same framework in Figure~\ref{fig:app_example} but varied in the choice of the acoustic model, acoustic features, waveform model, and training strategy. 
Only {\sysI} and {\sysII} used the inadequate MIDI spectrogram \cite{cooper2021midi}, while others used the refined version.
Systems with the ID \texttt{abs} are analysis-by-synthesis systems that used the acoustic features of natural audio as input to the waveform model. They were included to measure the waveform models' performances.

The investigated acoustic models included Transformer TTS \cite{li2019neural} using an implementation from ESPnet \cite{hayashi2020espnet}, and a variant of Tacotron \cite{wanv2017tacotron} using an implementation from \cite{yasuda2019investigation}. 
Following \cite{cooper2021midi}, we set the dropout rate of the postnet to 0.99 to alleviate the issue of exposure bias \cite{schmidt2019generalization}. We also set the reduction factor to $4$ and fixed the length of input to $800$ frames.  
For waveform models, we compared NSF (\texttt{nsf}) \cite{wang2019neural}, HiFi-GAN (\texttt{hfg}) \cite{kong2020hifi}, and NSF-GAN (\texttt{nsfg}), all of which were implemented on ESPNet.
NSF used the same configuration as that in \cite{cooper2021midi}, HiFi-GAN used the default configuration set by ESPNet, and NSF-GAN combined the two.
For a sanity check and fair comparison with \cite{cooper2021midi}, we used the original NSF implementation, which is not implemented in ESPnet\footnote{We manually extracted acoustic features and trained this waveform model using specific scripts. During synthesis, we dumped the acoustic model's output from ESPNet and drove the NSF to synthesize audio. }. This waveform model is denoted as \texttt{nsfs}.

The experimental systems were trained using either a separate-training or a joint-training strategy. In the former case, the acoustic and waveform models were trained separately. The Transformer acoustic model was trained using the Adam optimizer \cite{kingma2014adam} with a learning rate of $1$ for 200k steps, while the training strategy of Tacotron followed \cite{cooper2021midi}. 
The waveform models were trained using the AdamW optimizer with a learning rate of $2 \times 10^{-4}$ for one million training steps. 
For the joint-training strategy, each component was first trained using the same configuration as in the separate-training strategy, after which they were jointly fine-tuned with the Adam optimizer and a learning rate of $1.25 \times 10 ^{-5}$ for 200k steps.

\vspace{-1mm}
\subsection{Subjective evaluation}
\vspace{-1mm}

\begin{table}[!t]
\centering
\caption{Experimental systems and evaluation results. 
}
\vspace{1mm}
\label{tab:midi-to-audio}
\resizebox{\columnwidth}{!}{
\setlength{\tabcolsep}{1pt}
\begin{tabular}{lcccccccc}
\toprule
\multirow{2}{*}{System ID} & 
\multirow{2}{*}{\shortstack{Acoustic \\ model}} & 
\multirow{2}{*}{\shortstack{Acoustic \\ feature}} & 
\multirow{2}{*}{\shortstack{Wave. \\ model}} & 
\multirow{2}{*}{\shortstack{Joint \\ train}} &
\multicolumn{3}{c}{Obj. Eval.} & 
\multirow{2}{*}{\shortstack{MOS \\ (mean)}}\\
\cline{6-9}
                    & & & & & Pitch & Chroma & Spec. & \\
\midrule
Natural & - & - & - & - & - & - & - & 3.98 \\
\midrule
\multicolumn{2}{l}{Software-based baselines}  \\
\sysXIV & \multicolumn{4}{l}{Sample-based MIDI-to-audio s.w.} &  1.00 & 0.33 & 13.95 & 3.56 \\
\sysXV & \multicolumn{4}{l}{Physical-model MIDI-to-audio s.w.} & 0.92 & 0.32 & 12.16 & 4.10 \\
\midrule
\multicolumn{4}{l}{Synthesis system trained on flawed MIDI spectrogram}  \\
\sysI    & -    & midi-fb-f & NSF\cite{cooper2021midi} & - & 1.01 & 0.31 & 6.60 & 3.71 \\
\sysII   & taco & midi-fb-f & NSF\cite{cooper2021midi} & - & 1.18 & 0.37 & 9.65 & 2.95 \\
\midrule
\multicolumn{4}{l}{Waveform model trained on refined MIDI spectrogram}  \\
\absI   & - & midi-fb & NSF\cite{cooper2021midi}   & - & 1.31 & 0.38 & 5.72 & 3.31 \\
\absII  & - & midi-fb & NSF   & - & 1.37 & 0.39 & 7.20 & 3.35 \\
\absIII & - & midi-fb & NSF-GAN   & - & 1.26 & 0.34 & 5.14 & 3.69 \\
\absIV  & - & midi-fb & HiFi-GAN  & - & 1.16 & 0.31 & 4.69 & 3.80 \\
\midrule
\multicolumn{4}{l}{Acoustic model trained on refined MIDI spectrogram}  \\
\sysIII  & taco  & midi-fb & NSF\cite{cooper2021midi}  & - & 1.19 & 0.37 & 9.70  & 3.16 \\
\sysIV   & taco  & midi-fb & NSF  & - & 1.29 & 0.40 & 11.78 & 3.16 \\
\sysV    & taco  & midi-fb & NSF-GAN  & - & 1.11 & 0.35 & 9.09  & 3.18 \\
\sysVI   & taco  & midi-fb & HiFi-GAN & - & 1.58 & 0.56 & 10.07 & 2.21 \\
\sysVII  & trans & midi-fb & NSF\cite{cooper2021midi}  & - & 1.33 & 0.41 & 9.41 & 3.22 \\
\sysVIII & trans & midi-fb & NSF  & - & 1.42 & 0.44 & 10.94 & 3.10 \\
\sysIX   & trans & midi-fb & NSF-GAN  & - & 1.27 & 0.40 & 9.15  & 3.08 \\
\sysX    & trans & midi-fb & HiFi-GAN & - & 1.83 & 0.60 & 9.95  & 1.88 \\
\midrule
\multicolumn{4}{l}{Joint training of acoustic and waveform model}  \\
\sysXI   & trans & midi-fb & NSF  & \checkmark & 1.59 & 0.47 & 16.39 & 2.23 \\
\sysXII  & trans & midi-fb & NSF-GAN  & \checkmark & 1.12 & 0.38 & 9.09  & 3.32\\
\sysXIII & trans & midi-fb & HiFi-GAN & \checkmark & 1.10 & 0.38 & 9.14  & 3.58 \\
\bottomrule
\end{tabular}}
\vspace{-5mm}
\end{table}

\begin{figure}
    \begin{center}
    \includegraphics[scale=0.6]{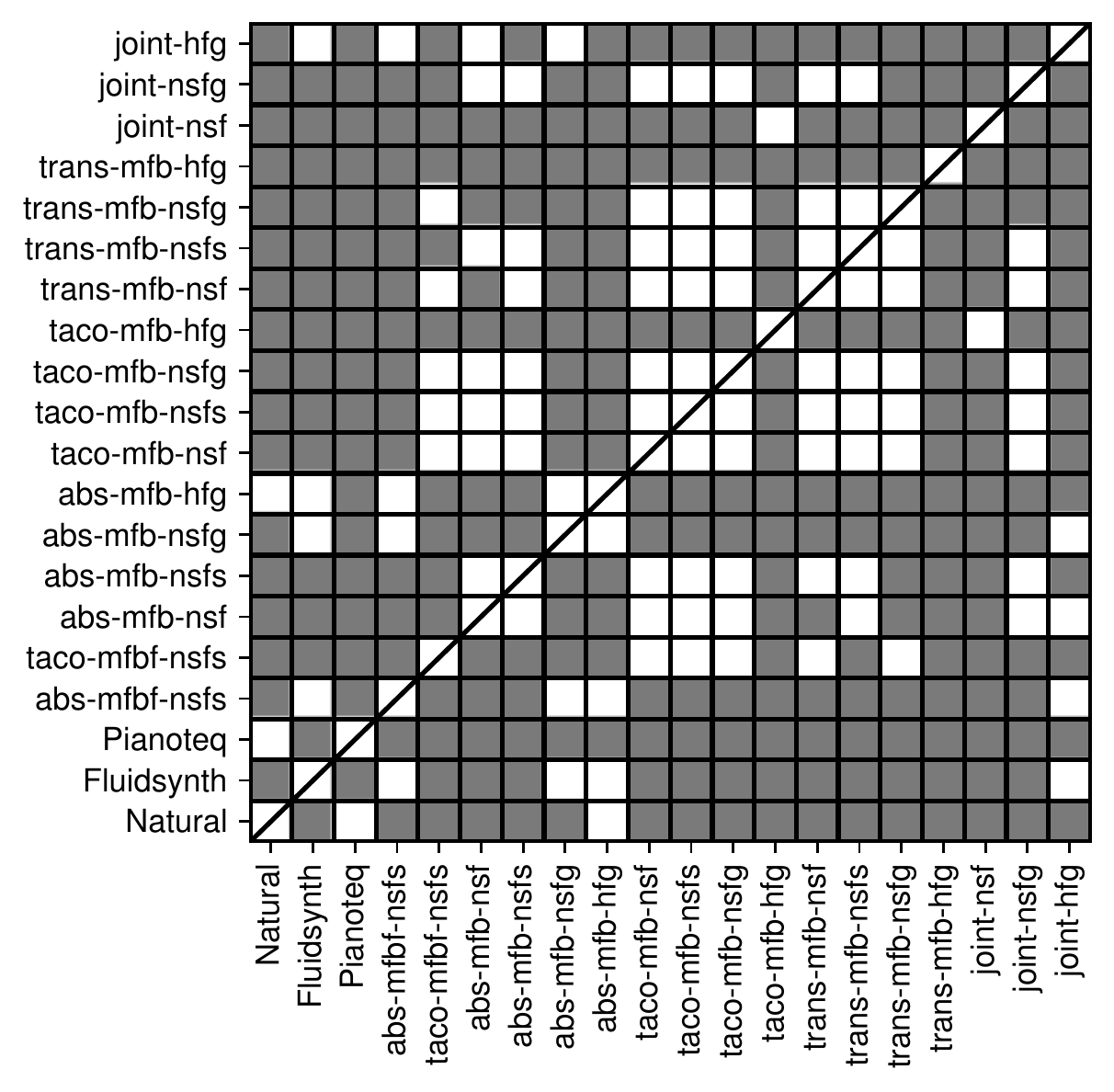}
    \vspace{-5mm}
    \caption{Results of two-sided Mann-Whitney U test with Holm-Bonferroni correction. Grey block indicates statistically significant difference at $\alpha = 0.05$}
    \label{fig:sig_test}
    \end{center}
    \vspace{-6mm}
\end{figure}

An extensive crowdsourced subjective listening test was conducted to evaluate the quality of the natural and synthesized audio from all the experimental systems.  
We split the audio samples on the basis of the locations of long pauses into segments that lasted from around 10 to 30 seconds. We then randomly selected 102 segment IDs and picked up corresponding segments from each systems as the listening test samples.  
Each listening test set had 30 different samples, balanced to contain at least one sample from each of the experimental systems.
Listeners were instructed to rate the overall quality of each sample using a scale from 1 (very bad) to 5 (very good), and they were allowed to rate at most 10 different sets. 
In total, there were 229 listeners, leading to 340 valid sets of results, and Table \ref{tab:midi-to-audio} shows MOS. 
We also conducted a statistical analysis using the two-sided Mann-Whitney U test with Holm-Bonferroni correction, the results of which are shown in Figure \ref{fig:sig_test}. The results allowed us to assess the effectiveness of the proposed refinements. 

\noindent\textbf{Did advanced acoustic and waveform models help?} In the analysis-by-synthesis condition, the HiFi-GAN waveform model {\absIV} achieved a MOS of 3.80, which outperformed the NSF models {\absI} and {\absII} with a statistically significant difference ($p\ll{1\mathrm{e}-10}$). Furthermore, {\absIV} was not statistically different from the natural recordings  
and the sample-based synthesizer {\sysXIV}. 
Using a GAN-based waveform model may be essential. In fact, by simply combining the HiFi-GAN's discriminators with NSF, the resulting NSF-GAN {\absIII} also outperformed the two original NSF models with a statistically significant margin. 

However, the gains brought by HiFi-GAN diminished when the acoustic features were predicted from separately trained acoustic models.  Systems {\sysVI} and {\sysX}, which used a HiFi-GAN waveform model, were significantly worse than the other systems using NSF-based models.  We argue that the reason is because the non-linear transformation in HiFi-GAN is sensitive to shifted parameters, while residual networks in NSF are able to encode correct synthesis parameters from distorted input.

The acoustic models did not show a significant difference in MOS when combined with a separately trained NSF-based acoustic model (\texttt{taco-mfb-nsf*} compared with each respective \texttt{trans-mfb-nsf*}). 

\noindent
\textbf{Did joint-training strategy improve system performance?} The results from the last six rows of Table~\ref{tab:midi-to-audio} showed that the joint-training strategy significantly improved the system performance if the system used a GAN-based waveform model. Particularly,   {\sysXIII}, which jointly optimized the Transformer and HiFi-GAN, achieved a MOS of 3.58, which was significantly higher than the counterpart {\sysX} using a separate-training strategy ($p\ll{1\mathrm{e}-100}$).  A similar improvement can be observed from the comparison between {\sysXII} and {\sysIX}. 

\begin{figure}[t!]
    \centering
    \includegraphics[scale=0.80]{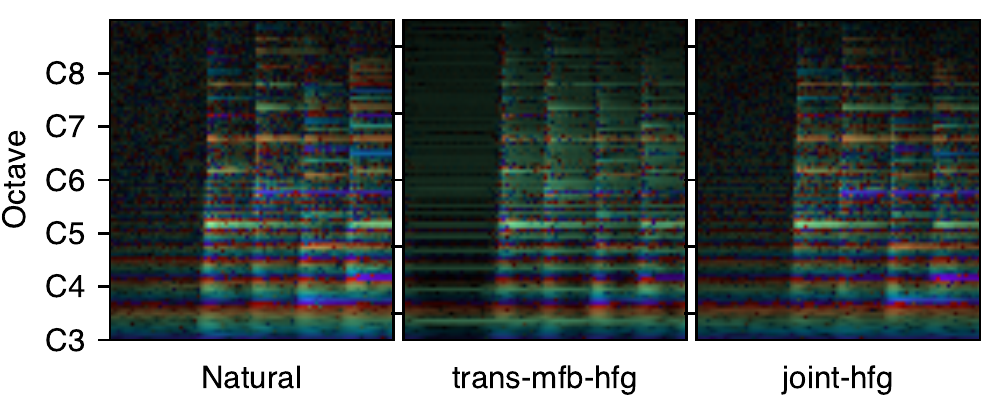}
    \vspace{-4mm}
    \caption{Rainbow-gram of Natural and HiFi-GAN-based systems.}
    \label{fig:maestro_rainbow}
    \vspace{-4mm}
\end{figure}

Figure~\ref{fig:maestro_rainbow} illustrates a Rainbow-gram\footnote{Rainbow-gram is based on STFT. It uses color and lightness to indicate the instantaneous frequency and spectral amplitude, respectively. When the spectral amplitude of a frequency bin is smaller, the image pixel is darker.} of natural and generated audio from {\sysX} and {\sysXIII}. From the natural audio's sub-figure, we can easily observe the notes and harmonics. In {\sysX}, the harmonics, especially in the high-frequency band, are less clear, and there was higher energy between harmonics. Furthermore, the low-frequency band has artifacts (e.g., horizontal bars) at the beginning of silence. Such degradation and artifacts were alleviated in the case of {\sysXIII}.

When using NSF, the joint-training strategy harmed {\sysXI}. One possible reason is that NSF is limited by maximum-likelihood-based STFT loss and lacks the capacity to produce high-quality waveforms from hidden acoustic features in a jointly trained system. Adding adversarial loss can help, and this suggests again the advantage of using a GAN-based waveform model.

\noindent
\textbf{Overall performance:} By combining the best components and training strategy, {\sysXIII} outperformed other MIDI-to-audio synthesis systems with statistically significant differences, including the system proposed in our previous work. Note that the refined MIDI spectrogram also contributed because the comparison between {\sysII} and {\sysIII} showed that the refined MIDI spectrogram was beneficial.
However, the MOS of {\sysXIII} was still lower than those of natural audio. This quality gap calls for further investigation, and part of it is presented in the objective analysis in the next section.

The physical-model-based {\sysXV} outperformed even the natural audio (although not statistically significantly), a finding consistent with previous work \cite{cooper2021midi}. However, the physical model is the outcome of laborious analysis and simulation, which does not easily generalize to other types of piano or different instruments.

\vspace{-1mm}
\subsection{Analysis with objective metrics}
\vspace{-1mm}

We measured a few distance metrics between the synthesized and natural audio, including the pitch, Chroma, and MIDI spectrogram distortion. The last two were computed using mean squared error (MSE). The 12-dimensional Chroma features were extracted using librosa \cite{mcfee2015librosa}. The pitch distortion was the cross-entropy between the estimated pitch probabilities of the synthesized and natural audio \cite{cooper2021midi}. The pitch estimation was done using  CREPE \cite{kim2018crepe}.

Results are listed in Table~\ref{tab:midi-to-audio}. 
We computed the correlation of these objective results with MOS. 
We found that both pitch and Chroma distortion had a negative correlation with MOS at $-0.88$ and $-0.91$, which indicates that the listeners were sensitive to out-of-tune notes.  The MIDI spectrogram distortion also showed a negative correlation with MOS on data-driven systems at $-0.68$.

On the basis of the above utterance-level analysis, we further analyzed how objective metrics correlate with the MOS at the note level. 
Given the MOS ratings for synthesized audio samples from one system,  we (approximately) computed an averaged MOS for each note that appeared\footnote{Since we asked the listeners to only give one rating to each audio sample in its entirety, we assigned the score to every note present in the audio sample. We accumulated the scores for each note and computed the mean.}. Since each dimension of the MIDI spectrogram corresponds to one note, the MIDI spectrogram distortion for each note was computed over the corresponding feature dimension. 
The note-level pitch distortion was computed using CREPE and cross-entropy without averaging over the notes.

\begin{figure}[t!]
    \centering
    \includegraphics[width=0.9\columnwidth]{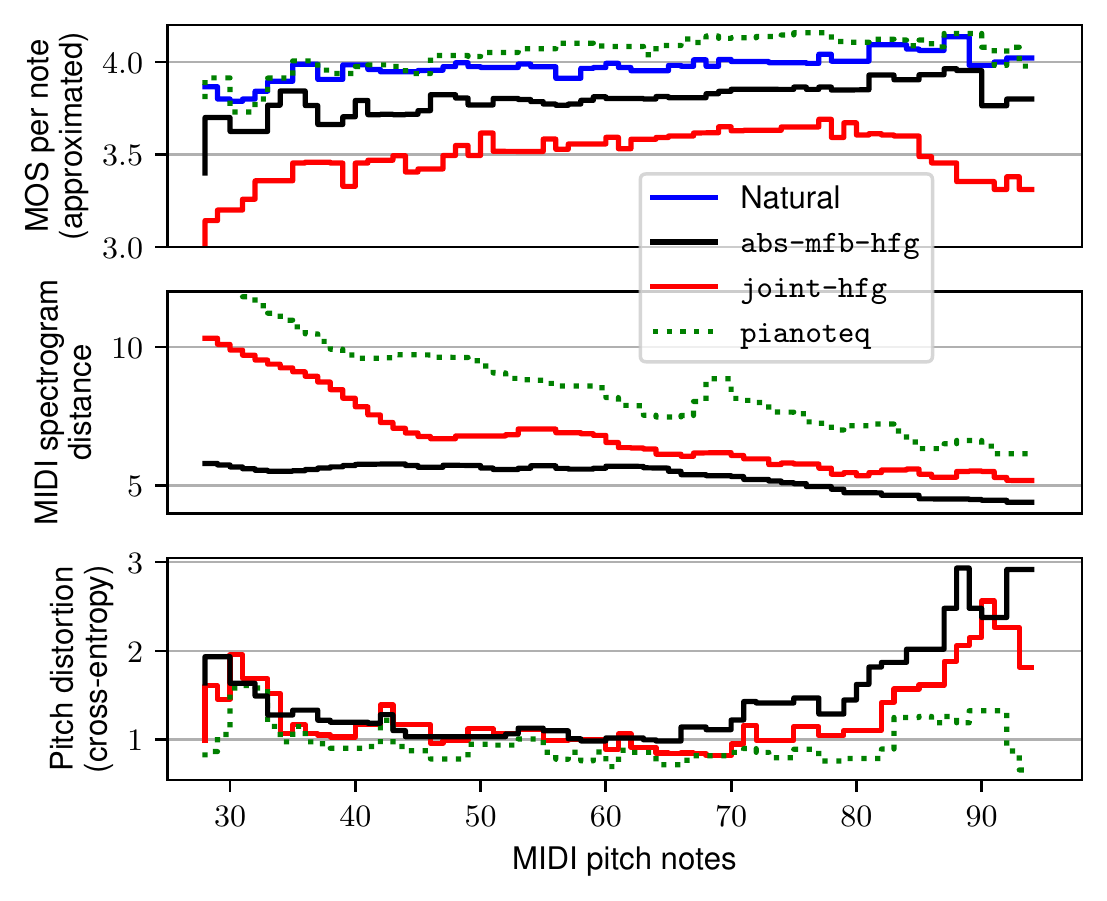}
    \vspace{-5mm}
    \caption{Approximated MOS per note in test set (top panel), pitch distortion (middle panel), and MIDI-spec. distortion (bottom panel).}
    \label{fig:testnote_pitch}
    \vspace{-4mm}
\end{figure}

Figure~\ref{fig:testnote_pitch} plots the results for {\sysXIII}, {\absIV}, and {\sysXV}. It can be seen that the MOS of {\sysXIII} was below {\absIV} and the natural audio for all the evaluated notes, and it degraded more on low- and high-pitched notes. 
The gap may be due to the distortion in the spectra of the generated audio. Particularly, the MIDI spectrogram distortion of {\sysXIII} on the low-pitched notes was much higher than that of {\absIV}. Note that {\sysXV}'s distortion was high because {\sysXV} is based on another piano rather than the one that played the natural audio. 
On pitch distortion, {\sysXIII} was better than {\absIV} on most of the notes, especially those larger than 70. This is consistent with the gross pitch distortion in Table~\ref{tab:midi-to-audio} and suggests that the gap between {\sysXIII} and {\absIV} may be mainly due to MIDI spectrogram distortion. 
However, notice that both {\sysXIII} and {\absIV} have a higher pitch distortion than {\sysXV}, especially on high-pitched notes. Compared with the physical-model-based {\sysXV}, the data-driven {\sysXIII} and {\absIV} need to further improve the pitch accuracy in order to boost the synthesized audio quality.
Due to the space limitation, more objective analysis on inadequate MIDI spectrogram systems is available on our \hyperref[projweb]{project webpage}.

\section{Conclusion}\label{sec:conclusion}

We synthesized high-fidelity piano audio from symbolic musical note input by modifying our previous TTS-based MIDI-to-audio system and introducing advanced TTS techniques.  Even though a performance gap still remains between the best MIDI-to-audio system and the physical model-based software, our proposed system is more advantageous in terms of efficiency and flexibility.  Through extensive analysis using subjective and objective evaluations, we measured the correlation between MOS and objective measurements, which indicates possibilities for further improvements in the future.
Considering that we are able to synthesize highly natural-sounding music with TTS-based techniques, we also expect to investigate more areas related to music synthesis, such as timbre transfer, multi-instrument audio synthesis, and performance generation in our future work.

\pretolerance=1000

\bibliographystyle{IEEEbib}
\bibliography{main}

\end{document}